# PRIVACY-PRESERVING ONLINE SHARING CHARGING PILE SCHEME WITH DIFFERENT NEEDS MATCHING

Zhiyu Huang[1,2]

[1]School of Computer Science and Engineering, Hunan University of Science and Technology, Xiangtan, China
[2]Hunan Key Laboratory for Service Computing and Novel Software Technology, Xiangtan, China

## ABSTRACT

*With the development of electric vehicles, more and more electric vehicles have difficulties in parking and charging. One of the reasons is that the number of charging piles is difficult to support the energy supply of electric vehicles, and a large number of private charging piles have a long idle time, so the energy supply problem of electric vehicles can be solved by sharing charging piles. The shared charging pile scheme uses Paillier encryption scheme and improved scheme to effectively protect user data. The scheme has homomorphism of addition and subtraction, and can process information without decryption. However, considering that different users have different needs, the matching is carried out after calculating the needs put forward by users. This scheme can effectively protect users' privacy and provide matching mechanisms with different requirements, so that users can better match the appropriate charging piles. The final result shows that its efficiency is better than the original Paillier scheme, and it can also meet the security requirements.*

## KEYWORDS

*Private charging pile sharing service, Privacy protection, Demand analysis, Homomorphic encryption, Internet of things*

## 1. INTRODUCTION

With the advancement of "carbon peak and carbon neutral" goals and the development of electric vehicles (EVs), EVs have the potential to effectively reduce air pollution caused by daily transportation [1]. As the number of EVs on the road increases, the demand for charging infrastructure also increases. However, the current prevalence and coverage of charging stations are insufficient to meet this demand[2-3]. Surveys have shown that between 2015 and 2020 in Table 1, the number of EVs and charging stations has been steadily increasing, with a particularly notable increase in the number of private charging stations, from 8,000 in 2015 to 874,000 in 2020[16]. However, compared to the growth rate of EVs, the number of charging stations is still far from adequate. As a result, for those EV users who are unable to install charging stations, the problem of charging difficulties is becoming increasingly apparent. In 2020, the Chinese government proposed to include charging stations as one of the fields of the nation's "new infrastructure", with an estimated investment of approximately 10 billion to build charging stations. According to international data surveys, it is expected that by 2030, there will be 5 million EVs on the road in California alone, and 12-24 million private charging stations and 10-20 million public charging stations globally. Charging facilities have become an indispensable infrastructure in new energy development planning [4-5]. Considering the high installation cost of charging piles [6], other technologies are needed to make up for the shortcomings of charging piles.

Table 1 Approximate number of electric vehicles and charging piles in the world

| Year | 2015 | 2016 | 2017 | 2018 | 2019 | 2020 |
|---|---|---|---|---|---|---|
| Number of electric vehicle | 570 | 1280 | 1840 | 2740 | 3890 | 4840 |
| Number of public charging piles | 58 | 149 | 240 | 387 | 516 | 807 |
| Number of private charging piles | 8 | 63 | 232 | 477 | 703 | 874 |

With the rapid development of Internet of Things technology, Internet of Things devices connect everything and have gradually entered the mode of Internet of Everything [7-9]. As one of the applications of the Internet of Things, the Internet of Vehicles can realize the information exchange between vehicles and provide certain research value and commercial value. The application of V2X technology of the Internet of Vehicles in cloud (edge) computing is the cornerstone of building a smart city and smart transportation [10-12]. At present, the research on shared charging pile is still in the initial stage. The traditional charging pile sharing scheme generally consists of three entities, including charging pile provider, electric vehicle and matching server, in which both buyers and sellers upload their own information to the server for matching calculation, and the server returns the matching results to both buyers and sellers, as shown in Fig 1. However, in the traditional charging pile sharing scheme, the user's information is published or uploaded to the server through simple encryption, and the server needs to decrypt the participants' information to get their plaintext information. Therefore, in the traditional scheme, the user's privacy may be attacked and leaked. In the traditional charging pile sharing system, all information will be published directly on the Internet. One of the biggest problems faced by the system is that users will expose their private information to the public platform when they apply for it. For example, a malicious user has used a certain charging station. The charging pile is marked and recorded, and he may not use it directly through the platform when he knows that the charging pile is unmanaged for a period of time.

At the same time, there is also the possibility that the shared service platform exposes the privacy of customers. Because the location information of electric vehicles may include workplaces, home addresses, special hospitals or frequently visited entertainment venues, buyers' hobbies and health status information are leaked, and the privacy of charging pile sellers will also be greatly affected threaten. On the other hand, once the information of buyers and sellers is obtained by malicious attackers, not only will there be profitable and targeted advertisements, but also related work and home addresses will be threatened, and may even lead to personal safety. Therefore, in order not to disclose the private information of customers, it is necessary to design a secure service platform. This paper proposes the use of homomorphic encryption technology to protect the privacy of users.

In order to meet the above challenges, the main contributions of this paper are summarized as follows:

1) We use homomorphic encryption technology to encrypt user information, and at the same time use homomorphic characteristics to process ciphertext, and match the obtained results in the cloud server. In the public service platform, users' effective information will not be exposed, and matching can be completed efficiently.

2) For users with different needs, we designed the demand parameter $\omega$. Through matching calculation, we can get the matching index parameter W. By comparing W, we can get the most suitable buyers and sellers. This requirement parameter can better match users with different requirements.

3) We use chinese remainder theorem (CRT) to speed up the modular exponentiation in the decryption process of cloud server, CRT is used to convert $a^b$ from $Z_{n^2}$ to $Z_{p^2}$ and $Z_{q^2}$ for calculation. We use Paillier scheme with optimized parameters, which can speed up the encryption calculation although it loses homomorphism.

The remainder of this paper is organized as follows: In Section 2, we introduce the homomorphic encryption, parameter optimization of Paillier scheme and china remainder theorem. In Section 3, we introduce the system model and present the proposed scheme. In Section 4, we describe the performance evaluation results. Finally, we conclude the paper in Section 5.

## 2. RELATED WORK

In this section, homomorphic encryption technology, parameter optimization of Paillier scheme and Chinese remainder theorem are introduced.

### 2.1. Homomorphic Encryption

Encryption technology is often used to protect privacy, among which homomorphic encryption is a special encryption method, which has the characteristics of directly calculating encrypted data, such as addition and multiplication, and will not reveal any information of the original text during the calculation process. And the calculated result is also encrypted, and the result obtained after decrypting the processed ciphertext with the key is exactly the result obtained after processing the original text. Paillier scheme has the homomorphism of addition/subtraction. For plaintext m1 and m2, there is a function E () that makes $E(m1+m2)=E(m1)\cdot E(m2)$. Paillier scheme satisfies the standard semantic security of encryption scheme[13], that is, the ciphertext is indistinguishable (IND-CPA) under the attack of selected plaintext, that is, the information about plaintext will not be leaked in ciphertext. Its security is proved by the hypothesis of deterministic composite residue. So far, no algorithm can be cracked in polynomial time, so Paillier encryption scheme is considered to be safe. The detailed process includes the following steps.

- **KeyGen() :** Pick two prime numbers p and q compute n = p * q and $\lambda$ =lcm(p-1,q-1). Choose a random number g, and gcd(L(g^$\lambda$ mod n^2),n) = 1, computer μ =(L(g^$\lambda$ mod n^2))^(-1) mod n, where L(x) = (x-1)/n the public and private keys are pk = (n,g) and sk =($\lambda$ , μ) , respectively.

- **Encrypt() :** Enter the plaintext message m and select the random number r. Encrypt plaintext:

$$c = g^m \cdot r^n \bmod n^2, \tag{1}$$

- **Decrypt() :** Enter ciphertext C. Calculate plaintext message:

$$m = L(c^\lambda \bmod n^2) \cdot \mu \bmod n \quad (2)$$

## 2.2. Parameter Optimization of Paillier scheme

In order to simplify computation without affecting the algorithm's correctness, the algorithm may take g=n+1 during the key generation phase[14]. This allows for the simplification of the calculation of $g^m$ during the encryption process.

For g^m=(n+1)^m, using the binomial theorem, we can express g^m as the sum of the product of the binomial coefficients and the corresponding powers of n and 1, where each term of the sum can be calculated efficiently.

$$(n+1)^m \bmod n^2 = \binom{m}{0} n^m + \binom{m}{1} n^{m-1} + \cdots + \binom{m}{m-2} n^2 + mn + 1 \bmod n^2, \quad (3)$$

As the previous m-1 terms are multiples of n, under the condition of modulo $n^2$ operation, they can all be eliminated, thus this modulo exponentiation operation can ultimately be simplified to one modulo multiplication operation, thus accelerating the encryption process.

$$c = (1+mn) \cdot r^n \bmod n^2, \quad (4)$$

Decrypt ciphertext c:

$$m = \frac{c^\kappa - 1 \bmod n^2}{n}. \quad (5)$$

## 2.3. Chinese remainder theorem

The Chinese Remainder Theorem, also known as the Sunzi Theorem, originates from the ancient Chinese mathematical treatise "Sunzi Suanjing" and describes the isomorphism of two algebraic spaces. Specifically, an algebraic space can be decomposed into several mutually orthogonal subspaces and the original space corresponds one-to-one to the decomposed space, similar to two forms of the same space. Specifically, when n = pq and p, q are relatively prime, there exists the algebraic isomorphism property: a mod n = a mod p + a mod q, thus the operations under mod n can be transformed into operations under mod p and mod q. By converting to this form, the calculation efficiency is higher. Therefore, this property can be utilized to accelerate modular exponentiation operations under mod n.

## 3. THE PROPOSED SCHEME

### 3.1. System Model

The matching scheme of shared charging piles consists of multiple electric vehicle buyers, multiple charging pile sellers, multiple edge proxy servers, a cloud server and a certificate certification center. All entities communicate through the mobile network. The Figure.1 describes our system model.

**Electric vehicle** (EVs): As a user of shared charging piles, when charging piles are needed, a charging request will be sent out, and the EV set is expressed as {1, ⋯, i, ⋯, I}. After receiving the response, you will get the public key of information encryption, and the terminal equipment of the Internet of Things will encrypt the information to be sent with the public key and send it to the nearest proxy server.

**Private charging piles** (PCPs): As a provider of shared charging piles, there will be J private charging piles in a given area, and the collection of charging piles is denoted as $\{1,\cdots, j\cdots, J\}$. Each PCP is managed by the owner and is equipped with a socket for EV charging. When each PCP has free time, it will issue an application for energy supply, and the provided information will be encrypted on the Internet of Things terminal equipment with the provided public key, and then the encrypted information will be sent to the nearest proxy server.

**Proxy server**: The proxy server has certain computing power, and is mainly responsible for collecting the encrypted information provided by nearby electric vehicle buyers and charging pile sellers who apply for matching, and using homomorphic characteristics to calculate the encrypted information. In the calculation process, important information is protected by Paillier, and the edge proxy server will not get useful information.

**Cloud server**: A cloud server is a server with powerful computing power, which can process encrypted information sent by proxy servers. After processing the information, use the matching scheme provided to match the buyer and the seller. After the matching, the best matching object will be obtained, and then the next round of matching will be carried out.

**Certificate certification** (CA): The certificate certification is the only authoritative identity certification institution and is completely reliable. All user entities need to be registered and authenticated by the certificate authority, and when the user sends an application, the corresponding public and private key pairs are generated by the key management center of the certificate authority and sent to the corresponding users. The information of the certificate certification center is absolutely confidential, and there is no possibility of collusion.

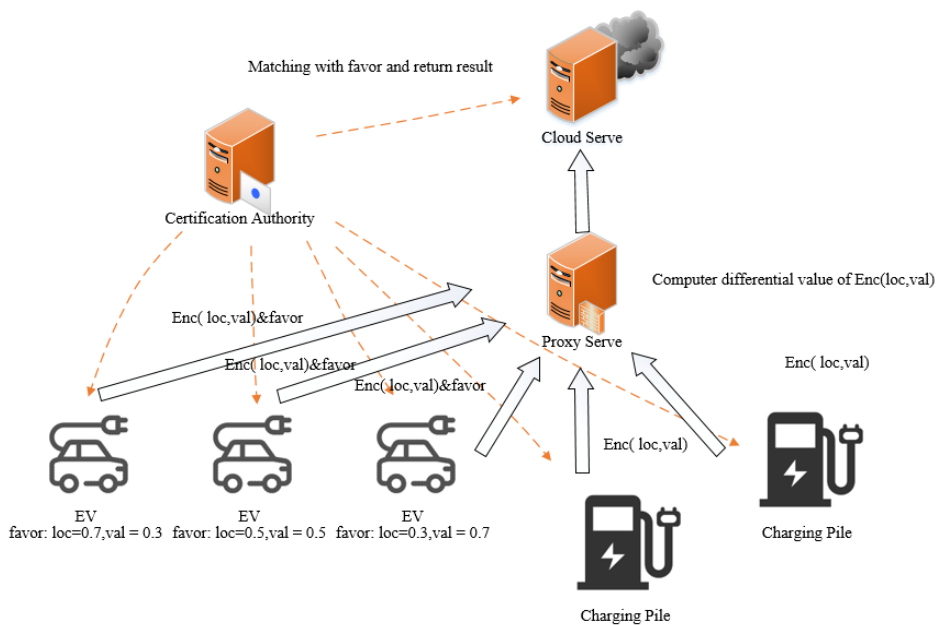

Figure.1 System model.

### 3.2. Requests and Information Encryption

In the shared charging pile matching system, the certificate certification center is responsible for managing and issuing public keys and maintaining public key information. In the system, electric vehicle buyers and charging pile sellers provide information, including location, price, demand and other information. Buyers of electric vehicles need to provide their own location $x_i, y_i$, proposed price ($r_i$), farthest acceptable distance $d_{imax}$, demand $\alpha_{i=1,2,\ldots,n}=\{0, 1\}$ and the price proposed by the demand $r_{\alpha i=1,2,\cdots,n}$. The seller of charging piles needs to provide the location $(x_j, y_j)$, the price ($r_j$), the demand $\alpha_{j=1,2,\ldots,n}=\{0, 1\}$ of charging piles and the price $r_{\alpha j=1,2,\cdots,n}$. After the user sends a request, the certificate certification center will send the public key pk to the user, and the electric vehicle buyer and the charging pile seller will select the random numbers $r_i$ and $r_j$ respectively, and use the public keys pk, $r_i$ and $r_j$ to encrypt the provided information.

$$C_{m_i} = \text{Enc}(m_i) = g^{m_i} \cdot r_i^n \bmod n^2, \tag{6}$$

$$C_{m_j} = \text{Enc}(m_j) = g^{m_j} \cdot r_j^n \bmod n^2, \tag{7}$$

After encryption, $C_m$ are obtained. The buyers of electric vehicles package $C_m$ and w and send them to the proxy server. The sellers of charging piles send $C_m$ to the proxy server. Except at the user end, the private information of users is all the information obtained after encryption and processing.

### 3.3. Ciphertext Processing

The proxy server has a certain computing power, and can use the homomorphic addition/subtraction characteristics of pailiier encryption scheme to process the ciphertext information as well as the plaintext. The process is as follows.

$$C_{a_{m_{ij}}} = C_{m_i} \cdot C_{m_j} = g^{m_i+m_j} \cdot (r_i \cdot r_j)^n \bmod n^2, \tag{8}$$

$$C_{d_{m_{ij}}} = C_{m_i}/C_{m_j} = g^{m_i-m_j} \cdot (r_i/r_j)^n \bmod n^2, \tag{9}$$

The demand of buyers and sellers is encrypted by homomorphic addition, and the information such as location, price and demand price is encrypted by homomorphic subtraction to get the sum of the processed information and the difference of the information. Among them, the main function of (8) is to judge whether the demand of electric vehicle buyers can be met. Information difference is a difference comparison between the information provided by buyers and sellers, which can reflect the information similarity of both parties. The smaller the information difference, the closer the information provided by the charging pile seller is to the preference of the electric vehicle buyer, which is more suitable for matching and has a higher matching probability. On the contrary, the larger the information difference between the two parties means that the user's matching probability will be smaller.

## 3.4. Information Decryption

The cloud server owns the private key sk issued by CA, including p and q. In Paillier cryptosystem, the main cost of decryption is modular exponentiation under $Z_{n^2}$. With the private key (decomposition p, q of n), the modular exponentiation under $Z_{n^2}$ can be converted into $Z_{p^2}$ and $Z_{q^2}$ by CRT.

The optimization function using CRT is expressed as $L_{p(x)}=(x-1)/p$ and $L_{q(x)}=(x-1)/q$ respectively, and the decryption process needs to be divided by using the following mathematical principles.

$$h_p = L_p(g^{p-1} \bmod p^2)^{-1} \bmod p \tag{10}$$

$$h_q = L_q(g^{q-1} \bmod q^2)^{-1} \bmod q \tag{11}$$

$$m_p = L_p(c^{p-1} \bmod p^2) h_p \bmod p \tag{12}$$

$$m_q = L_q(c^{q-1} \bmod q^2) h_q \bmod q \tag{13}$$

$$m = \mathrm{CRT}(m_p, m_q) \bmod pq \tag{14}$$

$\mathrm{CRT}(m_p, m_q \bmod pq)$ is to use CRT to calculate the modulus index, and the detailed process is as follows. For the modulus index $a^b \bmod n$, n=pq, CRT is used to convert $a^b$ from $Z_n$ to $Z_p$ and $Z_q$ for calculation. Calculate the mapping $a^b$ of $Z_p$ on $m_p = a_p{}^\wedge b_p$, where $a_p = a \bmod p$ can be obtained from euler theorem, where $\phi(p)=p-1$ is Euler function. Calculate the mapping $m_q = a_q{}^\wedge b_q$ of $a^q$ on $a^b$, which is the same as the calculation process of $m_p$. Calculate $m_p$ and $m_q$ separately and then aggregate them back.

$$m = m_p \cdot q^{-1}(\bmod\ p) \cdot q + m_q \cdot p^{-1}(\bmod\ q) \cdot p \tag{15}$$

Because p and q are coprime, there are $q^{-1}(\bmod\ p)q + p^{-1}(\bmod\ q)p = 1$. Substituting into the formula (15) gives:

$$m = m_p + (m_q - m_p)p^{-1}(\bmod\ q) \cdot p \tag{16}$$

In Paillier scheme, CRT is used to speed up decryption of plaintext to get plaintext m. After receiving the processed information, the modular operation under $Z_{n^2}$ is converted into modular operation under $Z_{p^2}$ and $Z_{q^2}$ by using private key sk by using China remainder theorem, and then decrypted.

$$\mathrm{amij} = \mathrm{Dec}(\mathrm{Camij}) = mi + mj \tag{17}$$

$$\mathrm{dmij} = \mathrm{Dec}(\mathrm{Cdmij}) = mi - mj \tag{18}$$

## 3.5. System Matching

After decryption by using CRT-optimized decryption scheme, the sum of information and the difference between information are obtained. After obtaining the decrypted information, first calculate the distance between the buyer i and the seller j:

$$d_{d_{ij}} = \sqrt{d_{x_{ij}}^2 + d_{x_{ij}}^2},\qquad(19)$$

The maximum acceptable distance of the EV buyer i is also encrypted and decrypted, because the maximum acceptable distance does not carry specific information such as location and price, so it is not regarded as important privacy information, so the cloud server gets the same clear text $d_{imax}$ as the user. To meet the matching conditions of buyer i, we must first compare the direct distance between buyer and seller. If $d_{dij}<d_{imax}$, it means that the distance between buyer i and seller j is less than the maximum distance accepted by buyer i, which meets the matching conditions. If $d_{dij}>d_{imax}$, the distance between users i and j does not meet the conditions, the seller j cannot match the buyer. Remove the unqualified sellers by comparison before proceeding to the next step.

Demand analysis is an interesting part of this paper. We consider that different buyers of electric vehicles may have different needs. Specifically, the distance and price are the information that the buyer i and the seller j must provide. Besides, other related demands α i and α j can be set, and the sum of them can be obtained through information and calculation. There are three situations:

Case1: $a_{\alpha ij}=0$, indicates that neither user has this requirement.

Case2: $a_{\alpha ij}=1$, it means that only one of the buyer i or the seller j owns the demand, and in case1 and case2, the corresponding i and j are removed because the demand cannot be provided.

Case3: $a_{\alpha ij}=2$, it means that i have the demand, and j can also provide the demand. At this time, whether to use it can be judged by the demand price difference $d_{rij}$.

i) when $d_{rij}<0$, it means that the price proposed by i is less than that proposed by j. At this time, i and j cannot match.ii)When $a_{\alpha ij}=2$ and $d_{rij}>0$ are met at the same time, it means that i and j meet the matching conditions, and the corresponding i and j are added to the matching set.

After getting the matching set that meets the distance and demand, the cloud server will make the final price matching. For buyer i, all sellers j who meet the demand will calculate the demand price difference $d_{rij}$ provided by i and j through formula (9). Because there is no information in the buyer's preferences that can lead to the leakage of location information, w are also unimportant information that can be obtained in the cloud server only after being decrypted by the private key sk. At this time, the information in the cloud server includes the location information difference $d_{dij}$, demand and value $a_{ij}$, price information difference $d_{rij}$, demand price difference $d_{rij}$, buyer's preferences w and the number k of sellers j in the number matching set. When matching the buyer i of the electric vehicle, the seller j in the matching set is calculated respectively to obtain $W_{ij}$.

$$W_{ij} = d_{dij} * w_{di} + d_{rij} * w_{ri} + \sum_{n=1}^{k} d_{\alpha_{ijn}} * w_{\alpha_{ijn}},\qquad(20)$$

For buyer i, $W_{ij}$ is used as a matching evaluation index to judge the suitability of matching with seller j. So we sort $W_{ij}$ in descending order. The smallest $W_{ijmin}$ is obtained, which means that the current sellers j and u are the most suitable matching objects in terms of price and demand, so i match j.

### 3.6. Matching Result Return

After the matching of buyer i is completed, the cloud server will send a request to the successfully matched i and j. User i and j use the Paillier scheme with optimized parameters to generate public keys $pk_i$ and $pk_j$ and send them to the cloud server. The cloud server generates a random number r, and encrypts the private key sk with $pk_i$ and $pk_j$.

$$C = Enc(sk)_{pk} = (1 + mn) \cdot r^n \bmod n^2, \qquad (21)$$

And the matched result is packaged and sent to the proxy server. The proxy server stores the encrypted address information uploaded by the user. After receiving the packaged result and the encrypted private key, the proxy server finds the corresponding encrypted address information $C_{Loci}$ and $C_{Locj}$ and the encrypted price information $C_{rj}$ of the seller j through the matching results i and j. The proxy server packages and sends the ciphertext of $C_{Locj}$, $C_{rj}$ and private key sk encrypted with public key $pk_i$ to buyer i, and packages and sends the ciphertext of $C_{Locj}$ and private key sk encrypted with public key $pk_j$ to seller j. The buyer I and the seller j use their own private keys to decrypt and get sk.

$$sk = Dec(C)_\kappa = \frac{c^\kappa - 1 \bmod n^2}{n}, \qquad (22)$$

Then use sk to decrypt the encrypted address information $C_{Loci}$, $C_{Locj}$ and $r_j$.

$$m = L(c^\lambda \bmod n^2) \cdot \mu \bmod n, \qquad (23)$$

At this time, the buyer I is matched, and the I+1th buyer is matched in the next round, and the matched seller J is eliminated from the matching set until the last seller set is empty, indicating that the current round of matching has ended. Re-apply to CA, get a new public and private key pair, and start the next round of matching.

## 4. PERFORMANCE EVALUATION RESULTS

### 4.1. Number Analysis

In this section, we consider a 3km*3km scene with different numbers of I buyers and J sellers. All the simulations are done on python and on 2.5 GHz Inter Core i5-7300HQ CPU and 32G RAM. Finally, all the simulation results are averaged in 50 simulations, and the consistent results are finally obtained.

Paillier encryption algorithm is a public key encryption algorithm based on number theory, which has high performance in security and time cost.
The following is the time cost of operating on a piece of data of the original Paillier encryption algorithm:

Randomly generate public key and private key: O(1)

Encryption operation: O(log n)

Decryption operation: O(log n)

Addition/subtraction operation (adding/subtracting two ciphertext numbers): O(1)

Where n refers to the length of the public key (the number of digits of the modulus).

In our simulation experiment, for I buyers and J buyers, the time cost from issuing an application to obtaining the corresponding public key is O(1). For all users, because the encryption operation uses the original Paillier encryption scheme, its time cost corresponds to O(log n). Our setting is that all users encrypt on the terminal equipment of the Internet of Things, and each user does it independently, so the time cost is fixed regardless of the number of matching users. When a user encrypts n data, the time cost is n*O(log n). When the terminal equipment of the Internet of Things encrypts the information to be sent, the proxy server will add/subtract the ciphertext data, and the corresponding operation is O(1). Every buyer I in the proxy server needs to add and subtract with the seller J. For J sellers and K pieces of information, the time cost is J*k*O(1). For I buyers, the total time cost required for calculation in the proxy server is I*J*k*O(1). Decrypt each calculated result in the cloud server. Under the condition that all sellers meet the requirements, the time cost of each decryption operation is O(log n) corresponding to I*J*k calculation results. After decryption, we execute the matching algorithm, calculate the matching index wij for M users who satisfy user I, and then get the minimum matching index wijmin for user I after sorting it, with the time cost of j*O(1). At this time, the buyer I and the seller J are successfully matched. The time cost corresponding to the above process is shown in Table 2.

Table 2 Paillier scheme time cost in this paper

|  | I buyers | J sellers | n data | Buyer matching | Index ranking |
|---|---|---|---|---|---|
| Encrypt | O(log n) | O(log n) | k*O(log n) | / | / |
| Process | / | / | k*O(1) | I*J*k*O(1) | j*O(1) |
| Decrypt | O(log n) | O(log n) | k*O(log n) | / | / |

In the process of decryption, we use CRT to speed up the calculation process. The original Paillier encryption algorithm needs to do modular exponential operation under $Z_{n^2}$. However, when the cloud server knows the private key sk and the corresponding coefficients p and q, the modular exponentiation under $Z_{n^2}$ is transformed to $Z_{p^2}$ and $Z_{q^2}$, thus improving the encryption and decryption efficiency. The time required to decrypt the ciphertext with the original Paillier encryption scheme and the time required to accelerate the calculation with CRT are shown in the figure. As can be seen from Figure.2, the decryption time is about 1/3 of that of Paillier encryption scheme after accelerated calculation with CRT. Compared with DJN scheme [15], the decryption time is basically the same as that of Paillier scheme. Therefore, using CRT to speed up the decryption process can effectively improve efficiency.

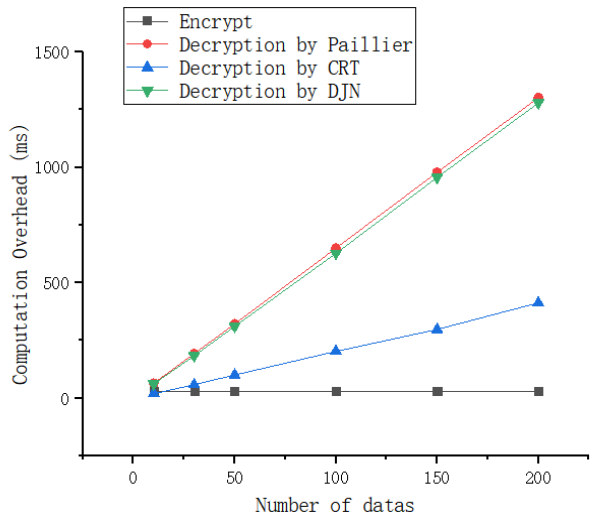

Figure.2 Computational overhead of data encryption and decryption with different schemes

After the cloud server gets the matching result, it sends out a successful matching application, and the buyer I and the seller J call the Paillier encryption algorithm with optimized parameters to generate a public key pair, and send the public key n to the cloud server, and the private key is stored at the user end. Compared with the original Paillier encryption algorithm, the encryption scheme with parameter optimization simplifies the modular exponential operation into a modular multiplication, which speeds up the encryption process. Its time efficiency is shown in Figure.3.

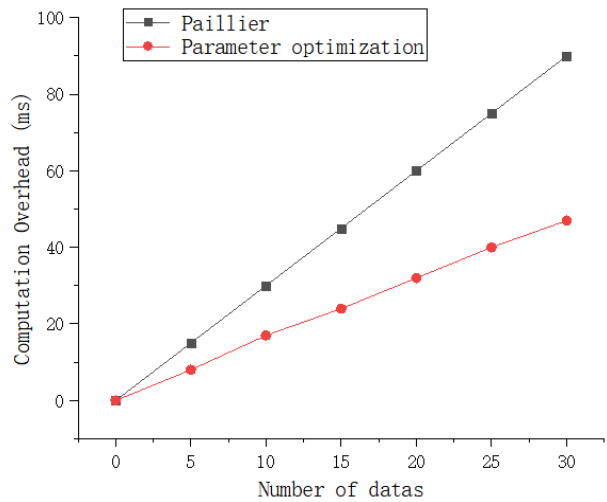

Figure.3 Comparison of encryption time cost between Paillier scheme and parameter optimization

## 4.2. Correctness Analysis

For ciphertext C in paillier scheme with optimized parameters, its correctness is expressed as:

$$\text{Dec}(C) = \frac{c^\kappa - 1 \bmod n^2}{n}$$

$$= \frac{((1+n)^m r^n)^{\tau\lambda} - 1 \bmod n^2}{n}$$
$$= \frac{(1+n)^{m\tau\lambda} - 1 \bmod n^2}{n}$$
$$= \frac{nm\tau\lambda \bmod n^2}{n} = \frac{nm}{n} = m, \quad (24)$$

For ciphertext C in Paillier scheme which uses CRT optimization for decryption, its correctness is expressed as:

$$\begin{aligned}
m &= m_p \cdot q^{-1}(\bmod\ p) \cdot q + m_q \cdot p^{-1}(\bmod\ q) \cdot p \\
&= m_p \cdot (1 - p^{-1}(\bmod\ q) \cdot p) + m_q \cdot p^{-1}(\bmod\ q) \cdot p \\
&= m_p + (m_q - m_p) \cdot p^{-1}(\bmod\ q) \cdot p,
\end{aligned} \quad (25)$$

Where $h_p$, $h_q$, $m_p$, $m_q$ and are obtained from formula (10), formula (11), formula (12) and formula (13) respectively.

### 4.3. Security Analysis

First of all, we assume that there are curious buyers and sellers, denoted as B, who want to attack through some information on the network to obtain other users' private information. In this paper, the original Paillier encryption scheme was adopted before the matching was completed. This scheme has been fully studied, so far there is no polynomial time algorithm to break it, so the security of Paillier encryption scheme is considered to be reliable. When B obtains the ciphertext message and the processed ciphertext message in the proxy server through attack, it cannot obtain effective information because there is no corresponding private key sk. So in the proxy server, we think the information is safe and reliable. When the processed information is sent to the cloud server, the cloud server needs to use the private key sk to decrypt the information. Suppose that B obtains the sum and difference of the decrypted information in the cloud server through special means attack, and these B have their own information, they will infer other useful information through the difference between the existing information and the information obtained by the attack in the cloud server. When inferring the position of other sellers through the difference between their own position information and the obtained position information, because there is only a straight distance, the inferred information cannot locate the specific position of the seller. From the above, even if the attack obtains the encrypted information in the proxy server or the decrypted information in the cloud server, B cannot infer the valid information. Therefore, for curious buyers and sellers, our scheme is safe and effective.

Secondly, for the premeditated attacker C, in our scheme, in order to prevent C from eavesdropping, all the communication between entities is encrypted. In our scheme, the random number r generated each time is different, and the encrypted result is also different. After attacking the information obtained by the proxy server and the cloud server, the user's location information and price information cannot be obtained through calculation. And we will refresh the key after each round of user matching. In this case, we think the scheme is also safe and effective.

## 5. CONCLUSIONS

In this paper, we solve the security problem of shared charging pile scheme through homomorphic encryption technology. In order to protect the privacy of users' location and provide matching strategies for users with different needs, we have formulated a privacy protection shared charging pile scheme based on users with different needs. First of all, we use the public key to encrypt the information in the terminal equipment of the Internet of Things, which effectively protects the privacy information such as location. Through homomorphism, the ciphertext matching the user is calculated in the proxy server, and CRT is used in the cloud server to accelerate the encryption process. We design the matching rules, calculate the matching index W and compare them to get the most suitable matching result. When we return the results, we use Paillier scheme with optimized parameters to effectively speed up the encryption process. Finally, our numerical analysis results show that the decryption time after CRT optimization is about 1/3 of the original Paillier scheme and DJN scheme. The encryption time after parameter optimization is 1/3 faster than that of the original Paillier scheme. At the same time, we also analyzed the security of the scheme, and the attacks of both curious users and malicious attackers are safe and reliable in the scheme on the public platform.

## REFERENCES

[1]  J. Zhang, H. Yan, N. Ding, J. Zhang, T. Li and S. Su, "Electric Vehicle Charging Network Development Characteristics and Policy Suggestions," 2018 International Symposium on Computer, Consumer and Control (IS3C), 2018, pp. 469-472.

[2]  S. Qiao, "Technical Analysis and Research on DC Charging Pile of Electric Vehicle," 2021 International Conference on Smart City and Green Energy (ICSCGE), 2021, pp. 89-93.

[3]  Y. Zhang, Y. Wang, F. Li, B. Wu, Y. -Y. Chiang and X. Zhang, "Efficient Deployment of Electric Vehicle Charging Infrastructure: Simultaneous Optimization of Charging Station Placement and Charging Pile Assignment," in IEEE Transactions on Intelligent Transportation Systems, vol. 22, no. 10, pp. 6654-6659, Oct. 2021.

[4]  A. J. Qarebagh, F. Sabahi and D. Nazarpour, "Optimized Scheduling for Solving Position Allocation Problem in Electric Vehicle Charging Stations," 2019 27th Iranian Conference on Electrical Engineering (ICEE), 2019, pp. 593-597.

[5]  H. Hu, S. Ni and L. Zhang, "Analysis of the carrying capacity of charging station based on regional charging demand," 2020 7th International Conference on Information Science and Control Engineering (ICISCE), 2020, pp. 1688-1691.

[6]  S. Mallapuram, N. Ngwum, F. Yuan, C. Lu and W. Yu, "Smart city: The state of the art, datasets, and evaluation platforms," 2017 IEEE/ACIS 16th International Conference on Computer and Information Science (ICIS), 2017, pp. 447-452.

[7]  I. M. Nafi, S. Tabassum, Q. R. Hassan and F. Abid, "Effect of Electric Vehicle Fast Charging Station on Residential Distribution Network in Bangladesh," 2021 5th International Conference on Electrical Engineering and Information Communication Technology (ICEEICT), 2021, pp. 1-5.

[8]  C. Liu, K. T. Chau, D. Wu and S. Gao, "Opportunities and Challenges of Vehicle-to-Home, Vehicle-to-Vehicle, and Vehicle-to-Grid Technologies," in Proceedings of the IEEE, vol. 101, no. 11, pp. 2409-2427, Nov. 2013.

[9]  Y. Wang, Z. Su and K. Zhang, "A Secure Private Charging Pile Sharing Scheme with Electric Vehicles in Energy Blockchain," 2019 18th IEEE International Conference On Trust, Security And Privacy In Computing And Communications/13th IEEE International Conference On Big Data Science And Engineering (TrustCom/BigDataSE), 2019, pp. 648-654.

[10] Zhao Tong,Feng Ye,Ming Yan,Hong Liu,Sunitha Basodi.A Survey on Algorithms for Intelligent Computing and Smart City Applications[J].Big Data Mining and Analytics,2021,4(03):155-172.

[11]  Anagnostopoulos T , Luo C , Ramson J , et al. A multi-agent system for distributed smartphone sensing cycling in smart cities[J]. Journal of Systems and Information Technology, 2020, ahead-of-print(ahead-of-print).

[12] Vidal S . Intelligent transport system in smart cities: aspects and challenges of vehicular networks and cloud[J]. Computing reviews, 2019(7):60.
[13] Paillier P . Public-Key Cryptosystems Based on Composite Degree Residuosity Classes[C]// Advances in Cryptology - EUROCRYPT '99, International Conference on the Theory and Application of Cryptographic Techniques, Prague, Czech Republic, May 2-6, 1999, Proceeding. Springer, Berlin, Heidelberg, 1999.
[14] Catalano D , Gennaro R , Howgrave-Graham N , et al. Paillier's Cryptosystem Revisited. 2002.
[15] Ivan, Damgrd, Mads, et al. A generalization of Paillier's public-key system with applications to electronic voting[J]. International Journal of Information Security, 2010, 9(6):371-385.
[16] https://baijiahao.baidu.com/s?id=1691272446545480912&wfr=spider&for=pc